\documentclass[%
 reprint,
 amsmath,amssymb,
 aps,
]{revtex4-2}

\usepackage{graphicx}
\usepackage{dcolumn}
\usepackage{bm}
\usepackage{amsmath}
\usepackage{latexsym}
\usepackage{textgreek}
\usepackage[version=4]{mhchem}

\usepackage[utf8]{inputenc}
\usepackage[T1]{fontenc}
\usepackage{mathptmx}
\usepackage{comment}
\usepackage{xcolor}
\usepackage{makecell}
\usepackage{textcomp}
\usepackage{hyperref}
\usepackage{mhchem}
\usepackage{siunitx}
\usepackage[switch,mathlines]{lineno}

\begin{document}

\preprint{APS/123-QED}

\title{Neutron Resonance Transmission Analysis with a Compact Deuterium-Tritium Neutron Generator}

\author{Ethan A. Klein}
\author{Farheen Naqvi}%
\author{Jacob E. Bickus}
\author{Hin Y. Lee}
\affiliation{Department of Nuclear Science and Engineering, MIT, Cambridge, MA 02139, USA}

\author{Robert J. Goldston}
\affiliation{%
Princeton Plasma Physics Laboratory, Princeton NJ 08543, USA
}%
\author{Areg Danagoulian}
\email{aregjan@mit.edu}
\affiliation{Department of Nuclear Science and Engineering, MIT, Cambridge, MA 02139, USA}%

\date{\today}

\begin{abstract}

Neutron Resonance Transmission Analysis (NRTA) is a spectroscopic technique which uses the resonant attenuation of epithermal neutrons to infer the isotopic composition of an object. NRTA is particularly well suited for applications requiring non-destructive analysis of objects containing mid- and high-Z elements. To date, NRTA has required large, expensive accelerator facilities to achieve precise neutron beams and has not been suitable for on-site applications. In this paper, we provide the first experimental demonstration that NRTA can be performed using a compact, low-cost deuterium-tritium (DT) neutron generator to analyze neutron resonances in the \SIrange{1}{50}{eV} range. The neutron transmission spectra for five single-element targets -- silver, cadmium, tungsten, indium, and depleted uranium – each show uniquely identifiable resonant attenuation dips in measurement times on the order of tens of minutes. Closely spaced resonances of $\sim$1 cm thick, multi-element targets can be easily differentiated with 1 eV resolution up to neutron energies of 10 eV and 5 eV resolution up to neutron energies of 30 eV. These results demonstrate the viability of compact NRTA measurements for isotopic identification and have the potential to significantly broaden the technique’s applicability across materials science, engineering, and nuclear security.
\end{abstract}

\maketitle


\section{\label{sec:level1}Introduction}

Most isotopes exhibit resonance behavior when interacting with neutrons. Some of the elements -- many of them with mid and high atomic number $Z$ -- undergo resonance scattering, absorption, and fission in the epithermal regime of 1-100 eV. In this energy range the resonances are sufficiently narrow and well separated that a given set of observed resonances constitutes a unique identifier of a particular isotope. Thus, spectroscopic measurement of neutron transmission and the resulting observation of attenuation dips in the transmitted spectrum can be used to infer the isotopic and elemental content of an unknown target. This is the basis of neutron resonance transmission analysis (NRTA).

In the past, NRTA has found applications in fields as broad as nuclear engineering, nuclear physics, and archaeology~\cite{ref:chichester2012JNMM,bourke2016non,schillebeeckx2015neutron,andreani2009novel,Sterbentz2010,Tremsin2013, Festa2015, hasemi2015evaluation,paradela2017neutron,C9JA00342H,Noguere2007a}. Furthermore, previous work in arms control has demonstrated the applicability of NRTA to problems of treaty verification~\cite{ref:hecla2018nuclear,ref:engel2019}. Using pulsed neutron beams and time-of-flight techniques, these applications have exploited the isotopic specificity of the NRTA signal. However, the  high neutron fluxes and long beam-lines necessary for sufficient spectral resolution have limited these applications to those that could be performed at facility-scale experimental setups. This circumstance has been a significant limitation to the broad applicability of an otherwise very powerful analytical technique. Recent work has sought to reduce the size of NRTA experimental setups by employing accelerator-based neutron sources, however even these involve fairly large experimental facilities~\cite{kusumawatix}. The prospect of using portable neutron generators and short beam-lines has been raised,~\cite{TSUCHIYA2018}  but no experimental results have yet been produced.

In this study, we provide the first experimental demonstration that NRTA can be performed using relatively cheap, compact, and commercially available pulsed DT neutron generators, thereby expanding the technique's potential applications. This effort requires a careful optimization of the moderator assembly and shielding to maximize the neutron output in the range of \SIrange{1}{100}{eV} and to reduce otherwise large neutron scatter and neutron capture backgrounds. Our past work via Monte Carlo (MC) simulations has indicated that such an approach is feasible~\cite{ref:engel2020}. NRTA applications using compact neutron sources can be further enhanced by applying the results from ongoing research into high-intensity, pulsed DT and deuterium-deuterium~(DD) neutron sources~\cite{ref:starfire,podpaly2018environment}.

\begin{figure*}[ht!]
  \centering
  \includegraphics[width=\textwidth]{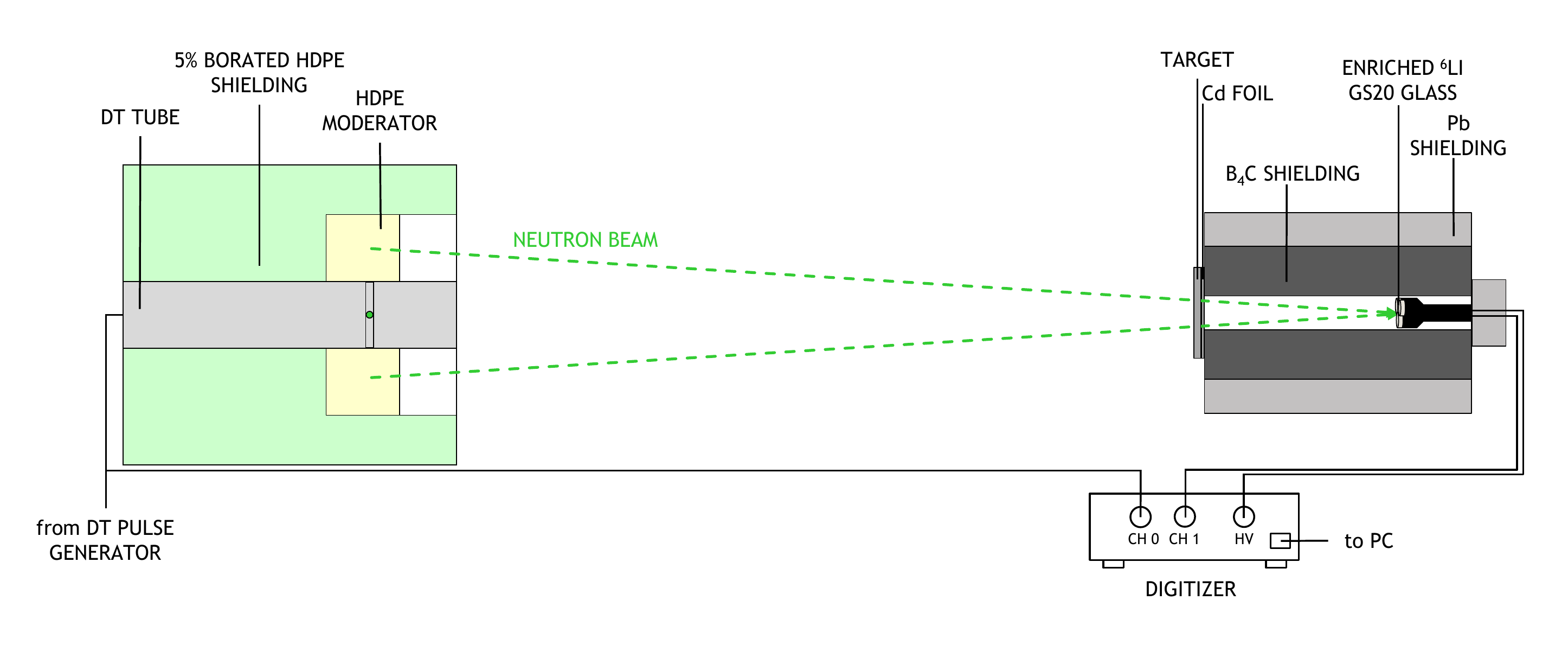}
  \caption{Schematic of the experimental setup. The DT neutron generator is shielded by a box of borated plastic with a radial moderator composed of high-density polyethylene. The moderated neutron beam is incident on a target and detected in a \SI{5}{mm} thick GS20 enriched $^{6}$Li glass scintillator coupled to a photomultiplier tube. The neutron generation signal and detector signal are both read out in a 14-bit, 250 MS/s digitizer. The detector is shielded from neutrons by boron carbide and from photons by lead. Thermal wraparound neutrons are filtered by \SI{3.0}{mm} of cadmium foil placed along the axis.}
  \label{fig:setup}
\end{figure*}

\section{\label{sec:background}Background}

The NRTA technique involves analysis of the magnitude and energy of attenuation dips in neutron transmission spectra to reconstruct the isotopic composition of a target object. In combination with evaluated neutron cross section data, the transmission spectrum can be used to estimate the linear density of the different isotopes present in the target object~\cite{hasemi2015evaluation}.\par
The total neutron cross section for a given material is a combination of both elastic potential scattering, which is nearly independent of energy, and resonant interactions, which are energy-dependent and specific to the isotopes present.
At the resonance energies, the enhanced cross section of forming a compound nucleus in a particular excited state can be approximated by the single-level Breit Wigner formalism~\cite{lynn1968theory}.
According to this formalism, the cross section for compound nuclear formation by entrance channel $\alpha$ and subsequent reaction by exit channel $\beta$, $\sigma_{\alpha\beta}$, is given by
\begin{equation}
  \sigma_{\alpha\beta} (E) \; = \; \frac{\pi}{k^{2}}\frac{\Gamma_{\alpha}\Gamma_{\beta}}{(E - E_{\text{res}})^2 + \left({\Gamma}/{2}\right)^2},
\end{equation}
where $\Gamma$ is the total decay width, $\Gamma_{\alpha}$ and $\Gamma_{\beta}$ are the partial widths for the entrance channel $\alpha$ and exit channel $\beta$, respectively, $k$ is the wavenumber, and $E_\text{res}$ is the resonant energy~\cite{wong1990introductory}. At energies of \SIrange{1}{100}{eV} the predominant resonant interaction is $(n, \gamma)$, which results in absorption of the incoming neutron and observation of a dip in the neutron transmission spectrum.

 Neutron transmission through a homogeneous target at a given incident energy, $T(E_{\text n})$, will depend on the identities and areal densities of the isotopes present in the target, and can be written as 
\begin{equation}
   T(E_{\text n}) = \frac{I_{}(E_{\text n})}{I_{0}(E_{\text n})} = \text{exp}\left( - \, \sum_{i} \, \sum_{j} \, n_{i} \, \sigma_{i,j}(E_{\text n}) \, \Delta x_{i}\right),
\label{transmission}
\end{equation}
where $I_{}(E_{\text{n}})$ and $I_{0}(E_{\text{n}})$ are the neutron fluxes measured at the detector for target-in and target-out setups, respectively, $n_{\text{i}}$ is the atomic number density for isotope $i$, $\Delta x_{\text{i}}$ is the target thickness for isotope $i$, and $\sigma_{\text{i,j}}(E_{\text{n}})$ is the interaction cross section for isotope $i$ and reaction $j$. This equation does not account for down-scattering and other secondary effects which may become significant for thicker targets. In the case where these effects are significant, MCNP simulations may provide a more accurate representation of neutron transmission for a given experimental setup. 

Since the resonance energies and amplitudes are characteristic of the isotopes present in the target, the neutron transmission spectrum will be unique for a given target isotopic composition that includes isotopes with epithermal resonances, making it an isotopic-geometric signature of a particular object, as described in Ref.~\cite{ref:engel2019}. Although some isotopes have individual resonances sufficiently close to those of another isotope such that the two cannot be distinguished, in most cases the combination of their multiple resonances can enable isotopic differentiation.

NRTA provides several benefits compared to gamma spectroscopy, including the ability to interrogate bulk high-Z material. NRTA can characterize any isotope which has sufficiently strong and well-resolved resonances in the neutron energy region of interest and can also provide geometric information by using a position-sensitive detector~\cite{Schooneveld2009,ref:tremsin,Tremsin2014}. This advantage is particularly important in arms control applications~\cite{ref:losko} and is supported by the fact that many isotopes of interest to nuclear security (\textit{e.g.}, \ce{^{235}U}, \ce{^{238}U}, \ce{^{239}Pu}, \ce{^{240}Pu}) exhibit (n,$\gamma)$ and (n,fission) neutron resonances in the epithermal energy region (\SIrange{1}{100}{eV}).
The capability of NRTA to perform isotopic identification and quantification will be limited by the thickness of the target and the presence of any shielding. 





\section{\label{sec:level2}Experimental Technique}

\subsection{\label{sec:setup}Experimental Setup}
The experiments were performed at the Vault Laboratory for Nuclear Science at MIT. A schematic diagram of the experimental setup is shown in Fig. \ref{fig:setup}. The neutron source was a portable A320 DT neutron generator, manufactured by Thermo Fisher Scientific Inc., that nominally generates $\sim 1 \times 10^{8}$ neutrons per second. The neutron generator was operated at a repetition rate of 5 kHz and a 5\% nominal duty factor with an acceleration voltage of 90~kV. Although the nominal pulse width is 10~\textmu s, due to voltage rise times and plasma formation time scales, the actual observed neutron pulse is only 3.3-\textmu s wide following an initial 7-\textmu s delay, as measured during experimental runs using the detector setup discussed in the next paragraph (see inlay of Fig. \ref{fig:TOF_spectrum}). These neutrons are produced nearly isotropically through the \ce{^{3}H}(\ce{^{2}H},n)\ce{^{4}He} fusion reaction, with neutron energies of 14.1 MeV in the direction of the moderator. To moderate the emitted neutrons to the epithermal energy range, a 10.0-cm thick, high-density polyethylene (HDPE) hollow, cylindrical moderator was placed radially around the DT tube. Surrounding the moderated source was 5\%-borated polyethylene shielding to reduce neutron and gamma backgrounds.\par

All experiments discussed in this paper used a distance of 2.60 m from the tritium target in the DT tube to the front face of the detector. The front face of the moderator was located at an axial distance of 5.0 cm from the tritium disk, corresponding to a time-of-flight (TOF) distance of \SI{2.55}{m}. MCNP simulations of the source-moderator assembly determined an effective moderation distance of \SI{2.0 \pm 0.1}{\cm}, which after adding to the true TOF distance, results in an effective TOF distance of \SI{2.57}{m}.

Target objects were thin, single-metal foils ranging from \SIrange{0.0127}{3.50}{mm} in thickness and of approximately 10 cm $\times$ 10 cm in area (see Table \ref{tab:table1}). For targets containing W, Ag, and In, foils of 3.5 mm, 0.25 mm, and 0.0127 mm thickness were used, respectively. For depleted uranium targets, individual 1.0-mm thick foils were stacked to provide the desired target thickness. For targets containing multiple elements, foils were stacked. The Cd foil was present for all sample runs (\textit{i.e.} target-in) and control runs with no target present  (\textit{i.e.} target-out), and was placed nearest to the detector (\textit{i.e.}, covering the collimator of the detector assembly.)

The detector setup consisted of three circular 25.4~mm $\times$ 5~mm disks of \ce{$^6$Li}-enriched GS20 scintillator glass~\cite{GS20} placed horizontally adjacent to each other, optically mounted on a 76-mm diameter, Hamamatsu R1307-01 photomultiplier tube (PMT) with a Scionix VD 14–E1 base and operated at 1150 V. Neutrons were detected in the GS20 glass via the \ce{^{6}Li}(n,\ce{^{3}H})\ce{^{4}He} reaction, depositing 4.78 MeV in the detector. This corresponds to the light output of approximately 1.6 MeVee~\cite{Oshima11}. The GS20 glass does not contain any elements possessing epithermal neutron resonances in sufficient quantities to compete with the dominant neutron detection mechanism. The 5-mm thickness of the scintillator disks was chosen to maximize neutron detection efficiency, while reducing the probability of gamma interaction and maximizing light collection. 

The detector and PMT were placed within an aluminum box filled with boron-carbide (\ce{B4C}) powder to shield thermal and epithermal neutrons, with a collimated opening in the front. The box was surrounded by 2"-thick, lead shielding to reduce gamma background. The initiation signal of the DT pulse and the PMT signal were read into two channels of a 14-bit, CAEN V1725 digitizer, which samples at the rate of 250 MS/s. Data acquisition was handled by the ADAQ toolkit~\cite{hartwig2016adaq} and data were analyzed using both ROOT~\cite{brun1997root} and the SciPy~\cite{2020SciPy-NMeth} and ImagingReso~\cite{zhang2017imagingreso} Python libraries.

\subsection{\label{sec:background}Background Reduction}
Accurately measuring the neutron transmission requires reducing the large backgrounds in the form of photons and unwanted neutron scattering within the detector assembly. The main source of $\gamma$ background is the \ce{^{1}H}(n,$\gamma$)\ce{^{2}H} reaction in the moderator surrounding the DT source, producing 2.2-MeV photons which are detected on the time-frame of neutron thermalization in the moderator (\textit{i.e.}, $\mathcal{O}$ (\SIrange{10}{100}{\micro s})). The digital signal analysis method of charge integration has been shown by others to improve neutron-gamma discrimination in PMT response from GS20 scintillators by a factor of over 60 compared to pulse height analysis~\cite{GS20ngammadiscrimination}, and so was chosen in this study to preferentially select neutron counts. As mentioned above, the thickness of the GS20 scintillator glass was chosen to be 5 mm to further reduce the photon contributions, while maintaining reasonable neutron detection efficiency (70\% at 1 eV, 30\% at 10 eV, 10\% at 100 eV). Another considerable background contribution to the detection of epithermal neutrons are the wraparound neutrons (\textit{i.e.}, slower neutrons incident on the detector during a subsequent pulse period). However, detection of these neutrons can be significantly reduced by the use of a cadmium filter, as will be discussed later.\par

Surrounding the detector with several centimeters of boron-carbide powder reduced the thermal-neutron background by almost 99\%. The impact of wraparound neutrons was minimized by a combination of a 200~\si{\us} pulse repetition period and the placement of 3.0-mm cadmium foil on-axis in front of the detector assembly. For the given pulse repetition rate and a distance of 2.60 m between the moderator assembly and the detector, only neutrons with energies $\sim$0.75~eV and below could result in wraparound. The (n,$\gamma$) cross section of $^{113}$Cd, a naturally occurring isotope, has a major, broad resonance at 0.17~eV of 10~kb which makes it an efficient material to absorb neutrons with energy below the $\sim$0.75~eV wraparound cutoff. The utility of NRTA is not affected much by the presence of the cadmium foil as very few isotopes have resonances below this cutoff (\textit{e.g.}, \ce{^{151}Eu}, \ce{^{191}Ir}). Furthermore, attenuation in the epithermal energy region due to 3.0-mm Cd is only $\sim$6\% and nearly independent of energy, and since Cd has no resonances below 65 eV (aside from a weak resonance at 27.6 eV), measurement of neutron resonances in the target is not impeded.

\subsection{\label{sec:TOF} Time-of-Flight (TOF) Measurement}
NRTA requires a means of measuring neutron energy in order to reconstruct the neutron transmission as a function of incident neutron energy. In these experiments, measurements of neutron time of flight were used to reconstruct neutron energy. For a defined TOF path length, $d_{\text{TOF}}$, the following non-relativistic, kinematic relationship can be used to infer the neutron energy, $E_\text{n}$:
\begin{equation}
 E_{\text{n}} = \frac{1}{2}m_{\text{n}}\left(\frac{d_{\text{TOF}}}{t_{\text{TOF}}}\right)^2
\label{TOF}
\end{equation}
where $m_\text{n}$ is the neutron mass and $t_\text{TOF}=t_{\text{signal}}-t_{\text{pulse}}$. Here $t_{\text{signal}}$ is the time of the detector trigger and $t_{\text{pulse}}$ is the center time of the neutron generation pulse. The resulting uncertainty in neutron energy is given by,
\begin{equation}
    \frac{\Delta \: E_n}{E_n}= 2\sqrt{\left(\frac{\;\Delta \: t_{\:\text{TOF}}}{t_{\text{TOF}}}\right)^2 + \left( \frac{\Delta \: d_{\:\text{TOF}}}{d_{\text{TOF}}}\right)^2 }
    \label{eq:uncertainty}
\end{equation}
where $d_{\:\text{TOF}}$ is the TOF distance, $\Delta \: d_{\:\text{TOF}}$ is the uncertainty in TOF distance, and $\Delta \: t_{\:\text{TOF}}$ is the uncertainty in TOF \cite{Brusegan2002TheExperiments}. Given the $\mathcal{O}$(\textmu s) pulse width of the DT neutron generation and chosen TOF distance, the energy uncertainty for this experimental setup is dominated by contributions from $\Delta \: t_\text{TOF}$ across the entire \SIrange{1}{100}{eV} range. Thus, neutron sources with narrower pulses would be necessary in order to access absorption lines at energies much higher than 50~eV. 

The uncertainty in TOF, $\Delta\:t_{\:\text{TOF}}$, has contributions from the DT pulse width and the time resolution of the detector, which for a $^{6}$Li glass scintillator is typically $\sim$60 ns \cite{GS20ngammadiscrimination}. MCNP simulations show that moderation time increases non-linearly as a function of final neutron energy with a corresponding non-linear uncertainty in moderation time. For this reason, uncertainty in moderation distribution was treated in terms of spatial uncertainty, rather than time uncertainty. The main source of time-of-flight uncertainty for this setup is thus due to the width of the DT pulse. For these experiments, the DT generator was set to produce a 10-\si{\us} pulse, the minimum pulse width possible for the A320 DT generator~\cite{DT_generator}. As discussed earlier, due to the plasma dynamics of neutron generation in the DT source, the actual width observed in neutron production was much shorter. The time distribution of neutron generation was measured by observing the 0-20 \si{\us} region of each NRTA experimental run and fitting a Gaussian to the peak in detector signal, and was found to be $\sigma=1.41\pm0.06$~\si{\us}. Experiments using an EJ-309 liquid scintillator capable of pulse-shape discrimination confirmed this peak to be a combination of neutron and gamma signals, and validated the time profile of neutron generation as measured using the GS20 glass.

The uncertainty in TOF distance is a result of the spatial spread of the neutrons following moderation from 14.1 MeV to epithermal energies. To minimize the impact of $\Delta \: d_{\:\text{TOF}}$ longer flight distances are preferred, but due to the reduction in neutron flux with distance, the need for reasonable measurement times sets a practical limit. A distance of 2.60 m, between the front face of the detector and the tritium disk within the DT tube, was thus chosen to support an acceptable TOF resolution while maintaining feasible measurement times. As mentioned above, for this experimental setup this distance corresponds to an effective $d_\text{TOF}$ of \SI{2.57}{\m} after accounting for neutron moderation. The flight time of $\sim$1~eV neutrons is on the order of \SI{185}{\micro s} for this distance, within the period of the DT neutron generator. Moderator studies using MCNP simulations found the standard deviation in moderation distance to be \SI{1.2}{\cm}, leading to an overall uncertainty in the TOF distance of approximately 1.5 cm. Given the pulse width mentioned above, this contributes an uncertainty in reconstructed neutron energy of < 0.1 eV in the epithermal range of interest and is not a limiting factor of spectral resolution.\par

According to equation \ref{eq:uncertainty}, the energy resolution for this experimental setup ($\Delta \: {t_\text{TOF}}$ = \SI{1.4}{\micro s} and $\Delta \: {d_\text{TOF}}$ = \SI{1.5}{cm}) is \SI{0.02}{eV} at \SI{1}{eV}, \SI{5.3}{eV} at \SI{50}{eV}, and \SI{15}{eV} at \SI{100}{eV}. 


\subsection{\label{sec:TOF} Data Acquisition \& Processing}
The data was acquired in a list mode, with one channel reading the signal pulses from the PMT coupled to the detector, while the other channel read the TTL pulses from the DT generator's driving circuitry. In the post-processing, each detector pulse was associated with a preceding TTL pulse and the time of flight was calculated. As both channels used ADCs driven by the same 250 MHz clock this allowed for determination of signal time with the electronic precision of \SI{4}{ns}. In addition to the TOF determination it was necessary to further filter the data to reduce the gamma backgrounds that were present due to the photon sensitivity of the GS20 scintillator. By integrating the pulse area, a quantity proportional to the light output of the scintillator was computed. The distribution of signal pulse areas was then analyzed to identify the main neutron peak due to the light output from alphas and tritons produced in the $^6$Li(n,$\alpha$)$^3$He reaction in the scintillator. A Gaussian fit was performed, and a $\pm 2\sigma$ cut was placed, thus significantly reducing the photon background. With the Cd filter in front and no target present, a data acquisition trigger rate of $<200$ Hz was observed in the $^6$Li detector.

\begin{figure*}[ht!]
  \centering
  \includegraphics[width=2.00\columnwidth]{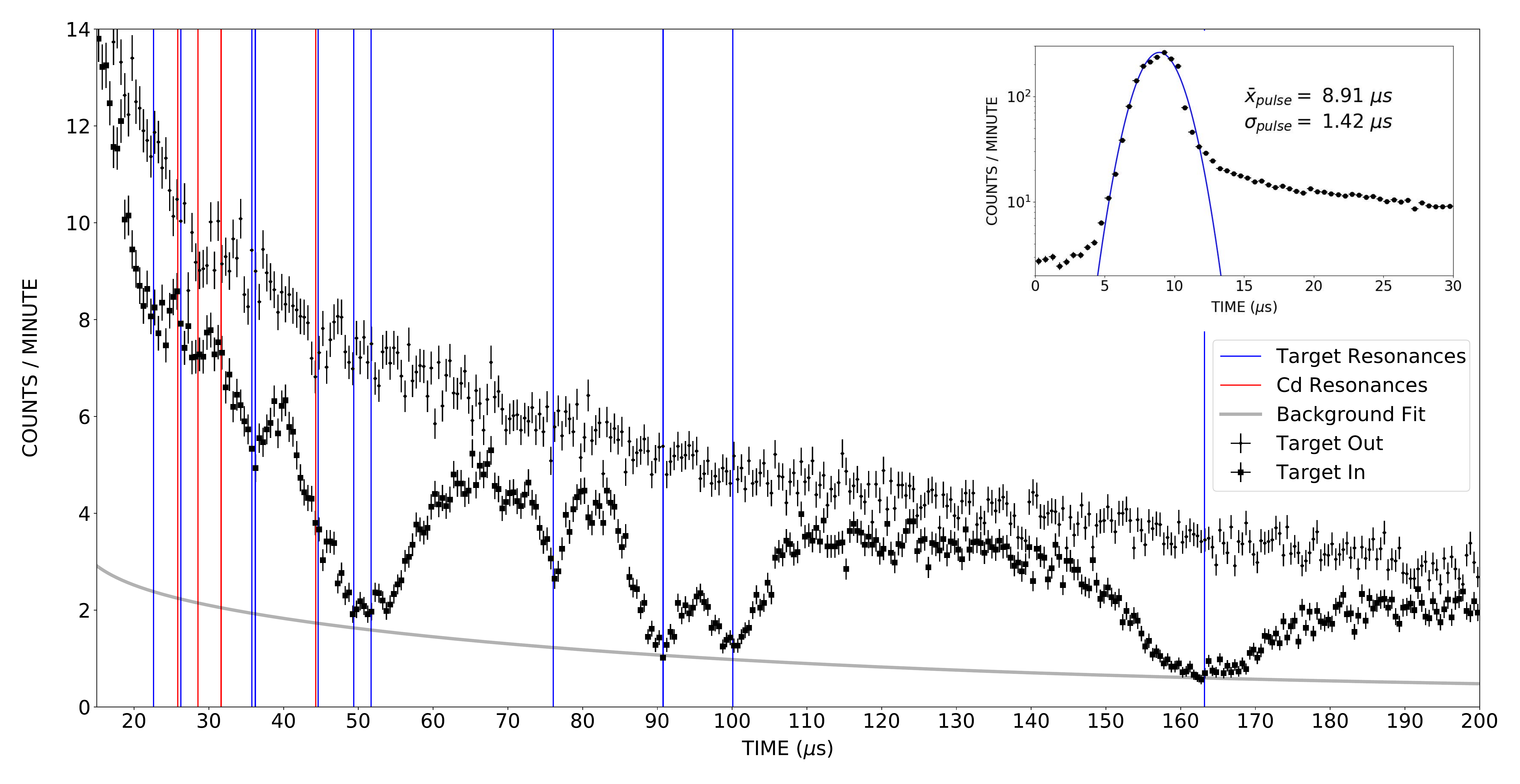}
  \caption{TOF distribution for a 2-hour run with a \SI{6.9}{mm} thick composite target containing W, Ag, Cd, and In. The background fit, shown in grey was performed according to the method discussed in Section \ref{sec:back_correction}. The location of the expected resonance energies for target isotopes (blue lines) and the Cd foil (red lines) matches the location of the observed attenuation lines. The inlay plots the total range of the neutron generation pulse, showing the DT pulse at $\sim$\SI{8.9}{\micro s} with a width of \SI{1.4}{\micro s}.}
  \label{fig:TOF_spectrum}
\end{figure*}

\begin{table}[h]
\caption{\label{tab:table1}%
Experimental Targets}
\begin{ruledtabular}
\begin{tabular}{ccc}
\textrm{Target\footnote{All target-in and target-out runs included the presence of a 3.0 mm Cd foil.}}&
\textrm{No. of Foils}&
\textrm{Foil Thickness (mm)}\\
\colrule
In, W, Ag & 3 & 0.127, 3.5, 0.25\\
In & 1 & 0.127\\
W & 1 & 3.5\\
Ag & 1 & 0.25\\
DU & 5 & 1.0, 1.0, 1.0, 1.0, 1.0\\
W, Ag, DU, Pb & 4 & 3.5, 0.25, 1.0, 1.1 \\
\end{tabular}
\end{ruledtabular}
\end{table}

\section{\label{sec:Analysis}Results \& Analysis}



Experimental measurements ranging from 10-minute to 2-hour runs were performed for various target compositions of W, Ag, In, and depleted uranium (99.9\% \ce{^{238}U}) ranging in thickness from 0.25 mm to 3.5 mm (see Table \ref{tab:table1}). The raw counts within the neutron cut described in the previous section were normalized to counts per minute and binned with \SI{0.5}{\micro s} bin width. Fig.~\ref{fig:TOF_spectrum} shows an example plot of a target consisting of In, W, and Ag in addition to the Cd-filter foil, along with the corresponding resonance energies from ENDF/B-VIII.0 libraries plotted as vertical lines~\cite{BROWN20181}. Distinct absorption lines for indium, tungsten, and silver isotopes are clearly observable, as well as Cd lines in the target-out spectrum, allowing for subsequent element identification. As part of future work, the absolute concentrations of the isotopes can be extracted from this data --- a goal that is particularly important for arms-control applications where the enrichment of a fissile material is of importance.

\begin{figure*}[ht!]
  \centering
  \includegraphics[width=1.95\columnwidth]{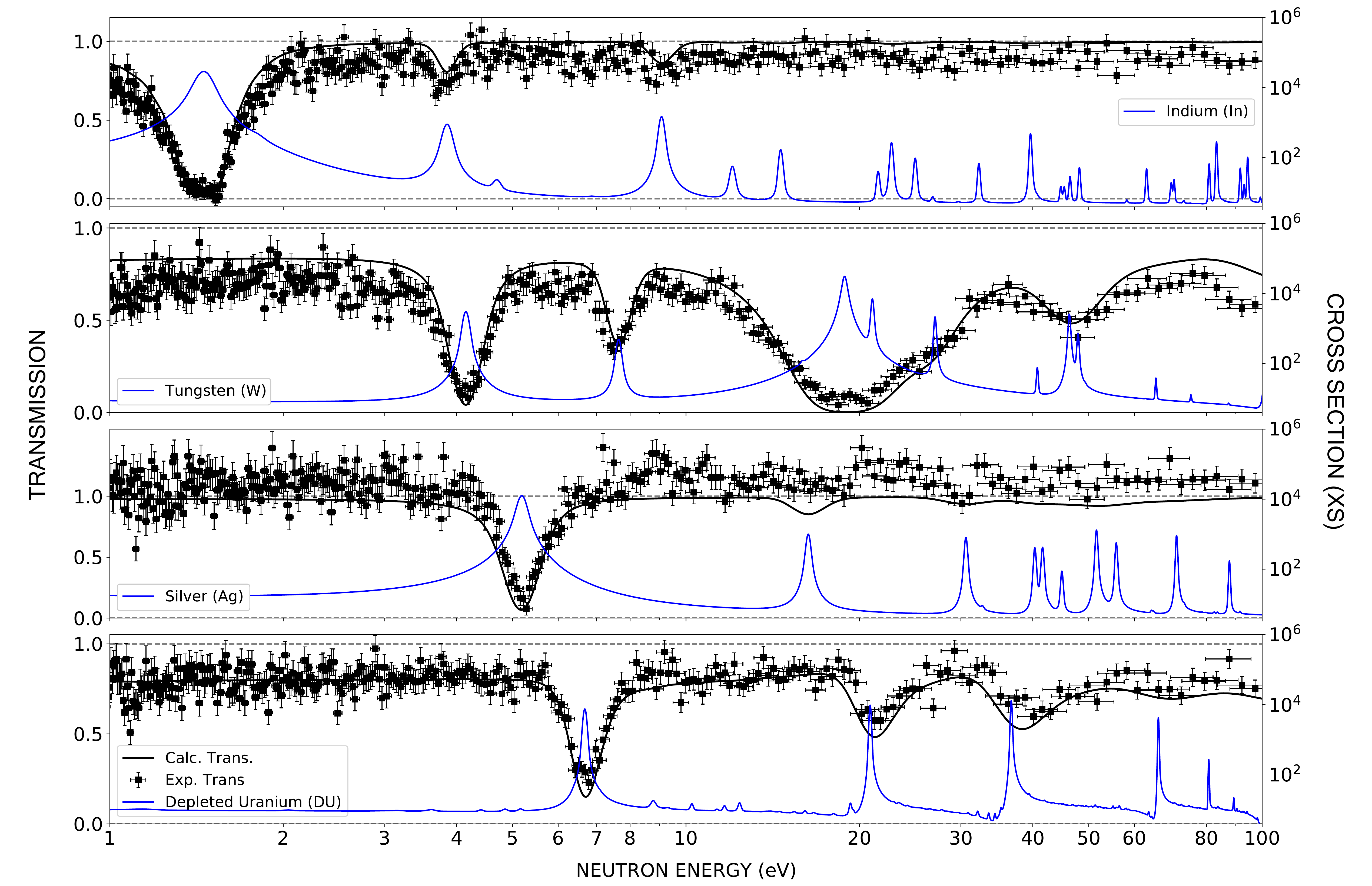}
  \caption{Neutron cross section and transmission data for single element targets (Top-to-bottom: In, W, Ag, depleted uranium). Measurement times were two hours for all runs with the exception of Ag, which was a one-hour measurement. Experimental data is plotted along with the calculated transmission, which is a convolution of ENDF/B-VIII.0 evaluated neutron cross-section data with the pulse width of the DT neutron generation. Horizontal dotted lines demarcate zero transmission (T=0) and full transmission (T=1).}
  \label{fig:elements}
\end{figure*}

\subsection{\label{sec:back_correction} Background Correction}
The background reduction methods described in \ref{sec:background}, in addition to the detector pulse integral cut, eliminated most of the unwanted photon and scattered neutron counts from the TOF spectrum. However, some residual background was present in the data, as observed at the bottoms of attenuation dips which should have zero counts due to very high neutron cross sections. Therefore, an additional correction function was applied to the data to estimate and subtract the residual background.

The background was modeled using the "black resonance" technique where a strongly resonant target (\textit{e.g.}, tungsten, silver, indium) is placed in the beam path and the background function is fit to the bottom of the “saturated” attenuation dips. This technique is described in detail by Syme et al. in Ref.~\cite{Syme1982}. A function of the following form was used following Fei \textit{et al.}~\cite{C9JA00342H}:
\begin{equation}
  B(t_{m}) = b_{0} + b_{1}e^{-t_{m}/\tau_{1}} + b_{2}t_{m}^{-\tau_{2}}
  \label{bg_fit}
\end{equation}
where $t_{m}$ is the neutron detection time corrected for the median pulse time, and $b_{0}$, $b_{1}$, $b_{2}$, $\tau_{1}$, and $\tau_{2}$ are fitting parameters. The first term in Eq.~\ref{bg_fit} accounts for constant background, the second accounts for the TOF distribution of 2.2-MeV $\gamma$ from neutron capture on $^{1}$H, and the third accounts for the background due to neutrons that underwent elastic scattering within the detector assembly before detection.\par

This function was fit to the resonance lines of a black resonance target containing 0.127-mm In (1.46 eV), 3.5-mm W (4.15 eV, 18.84 eV), and 0.25-mm Ag (5.19 eV) measured during a dedicated 2-hour run. To quantify the contributions of $2.2$-MeV photons, separate dedicated experiments were performed involving target-in and target-out runs using a \ce{LaBr3} detector. A decay constant of $\tau_{1}=$ $75\pm3$~\si{\us} was extracted using the \ce{LaBr3} data, and was used as a bound to the fit described by Eq.~\ref{bg_fit}. With $\tau_1$ thus determined, the neutron data was then fitted with Eq.~\ref{bg_fit} to determine the other parameters.\par

In applying the fit, attenuation of the background component from 2.2-MeV photons in the target must be taken into account. Photon attenuation in each target was calculated using the number densities and mass-attenuation coefficients~\cite{NIST} for each element present in the target and the ratio was taken to the attenuation in the "black resonance" target used for the background fit. This correction was not applied to the neutron component of the target under the assumption that neutron scattering within the detector assembly would be independent of target thickness, which was evidenced by statistical analysis of the best fit. Fig. \ref{fig:TOF_spectrum} shows the obtained fit function providing an estimate of the energy-dependent background counts that were then subtracted from the experimental data.\par

\begin{figure*}[ht]
  \includegraphics[width=2.00\columnwidth]{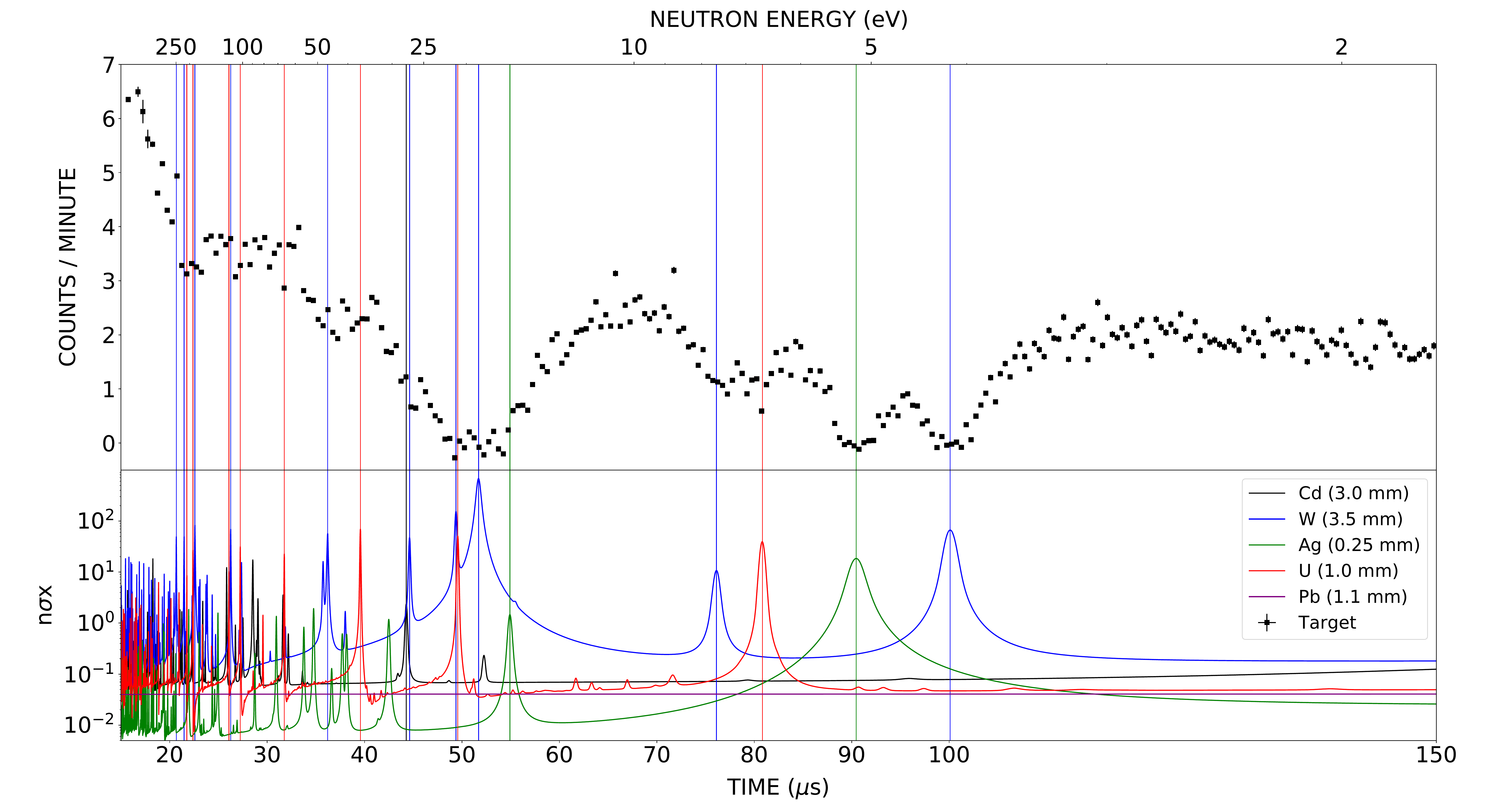}
  \caption{Background-corrected transmitted neutron counts for a multi-element target. The data was collected during a 2-hour run with an 8.9-mm thick target containing W, Ag, U, Pb, and Cd. The product of each element's total neutron cross section with its number density and elemental thickness is plotted beneath. At low incident neutron energies ( < \SI{10}{eV}), resonances within \SI{1.0}{eV} are well resolved.}
  \label{fig:DU_W_Ag}
\end{figure*}

\subsection{\label{sec:NRTA-elements} NRTA of Single-Element Targets}
Targets composed of a single mid- or high-Z element with strong neutron resonances were analyzed to measure the capability of the system to resolve resonances of varying widths across the epithermal energy range. The TOF histograms for target-in and target-out measurements were first background-corrected using the process described above, and the target-out spectrum was filtered. Then neutron transmission through the target was calculated by dividing the target-in histogram by the target-out histogram. Fig.~\ref{fig:elements} shows the background-corrected neutron transmission for single-element targets and plots the elemental neutron cross section for comparison. The transmission was also computed by convolution of ENDF/B-VIII.0 cross-section data with the shape of the DT neutron pulse, and is overlaid as the black line in the plot~\cite{BROWN20181}. 

Neutron resonances of sufficient width can be identified up to 50 eV. Although this upper energy bound is much lower than that of fixed-facility NRTA setups, which can identify resonances up to the keV range, the neutron transmission through each element presented here is clearly identifiable and matches the calculated transmission, demonstrating sufficiency for some isotopic-identification applications. For nuclear-security and safeguards applications, $^{238}$U resonances at 6.7 eV, 20.9 eV, and 36.7 eV are well resolved and thus can be used to indicate the presence of \ce{^{238}U} in the target. Although the presence of $^{238}$U can also be measured by gamma spectroscopy using the 1001-keV gamma emitted from $^{234m}$Pa in the $^{238}$U decay chain, determination of uranium enrichment can be challenging for thick or shielded objects. Gamma spectroscopy has been used for decades to accurately determine \ce{^{235}U} enrichment in samples~\cite{greenberg1987high}, but the technique relies on measuring low-energy gammas (usually the 185.7 keV from \ce{^{235}U} decay), which due to their attenuation in high-Z materials, limits the interrogation to the surface of the target and not the bulk~\cite{reilly1991passive}. In contrast, neutrons interrogate the entire target and the resulting transmission spectrum is a product of interactions throughout the target, not just the surface. 

The value of the reconstructed transmission is heavily dependent on the choice of background-fit parameters and target-out spectrum. While achieving a good overall agreement, it can also be seen to diverge from the calculated transmission in non-resonant regions is in some cases by approximately 10\%. This will not have an effect on elemental detection and attribution as that goal primarily depends on the observation of resonant dips. This however could affect quantification of elemental concentration in the target as it relies on the magnitude of the attenuation in the dip to that between the resonances to determine isotopic number densities. Thus future work should focus on better understanding and modeling of the backgrounds, \textit{e.g.}, via detailed Monte Carlo simulations, which would allow for a more precise determination of the transmission and more accurate estimations of isotopic concentrations in the target.

\subsection{\label{sec:NRTA-elements} NRTA of Composite Targets}
Isotopic identification of multi-element targets requires the ability to differentiate individual resonances. Fig.~\ref{fig:DU_W_Ag} shows the background-subtracted transmission for a 2-hour run with a composite target of W, Ag, Pb, and depleted uranium foils (in addition to the fixed Cd foil). As can be seen from comparison to the elemental neutron cross sections, all resonances below 20 eV can be attributed to individual elements known to be present, whereas higher-energy resonances are observed but due to their close spacing are not always attributable to specific elements.

The spacing between the 20.9-eV resonance of \ce{^{238}U} and the 18.8-eV \ce{^{186}W} and 21.1-eV \ce{^{182}W} resonances is too close to resolve with the 3.3-$\mu$s wide pulse of the DT generator. For that reason, nuclear-security applications requiring identification of $^{238}$U in the presence of tungsten will have to rely on measuring the 6.7-eV \ce{^{238}U} resonance , unless narrower DT pulses are used (\textit{e.g.} those achievable by a P235 DT portable neutron generator~\cite{ref:p325}) or a longer TOF distance is used with a corresponding increase in measurement time. If a longer TOF distance is chosen, it must be selected such that the TOF of the lowest energy neutrons not attenuated by the Cd foil is smaller than the neutron-pulse period, in order to prevent wraparound neutrons. Similarly, given fixed constraints on duty cycles of portable neutron generators, a shorter neutron-generation pulse may correspond to a shorter pulse period, thus requiring a shorter TOF distance or a thicker Cd filter to prevent wraparound neutrons.

\section{Conclusion}

In this work we have demonstrated the feasibility of performing isotopic identification in under an hour with NRTA measurements using a compact DT generator and short TOF distances. The measurements showed the resonant (n,n) elastic scattering and (n,$\gamma$) attenuation dips in the \SIrange{1}{50}{eV} range from medium- and high-$Z$ elements. This compact and economic configuration can significantly increase the applicability of a powerful technique which in the past has been extremely limited due to the necessity of large accelerators and long neutron beam-lines. This in turn would allow for practical applications in such fields as materials analysis, archaeology, and nuclear engineering. Verification of arms-control treaties in particular would benefit significantly from the use of such compact platforms in combination with already-proven NRTA methods, as described in Ref.~\cite{ref:engel2019,ref:hecla2018nuclear}. In addition, more intense neutron sources than the type used in this study may enable analysis of plutonium and fission products in spent nuclear fuel\cite{ref:chichester2012JNMM}, another scenario where a compact NRTA setup for on-site analysis would be needed to limit the transport of special nuclear material.

In its current form the technique does have a number of limitations and would benefit from the following future improvements. The spectral resolution is limited by the 3.3-$\mu$s neutron-generation pulse, which reduces the ability to resolve absorption lines above \SI{50}{eV}. The pulse can be shortened either by modifying the DT operation, or using more advanced, faster pulsed neutron sources such as those described in Ref.~\cite{podpaly2018environment,ref:p325}. This is of particular importance when it comes to elements whose powerful resonances below \SI{50}{eV} result in the saturation of the absorption lines: to quantify the isotopic concentrations one would need to observe and quantify the weaker, higher energy resonances. The technique is also limited by the presence of residual photon backgrounds from (n,$\gamma$) neutron capture inside the moderator. While energy-based filtering and background corrections have been effective, the sensitivity could be further improved by using detectors with a higher ability to discriminate between neutron and photon counts. For example, CLYC detectors appear promising in this regard~\cite{bourne2014characterization,dolympia2013pulse}. Finally, these measurements would benefit from a more powerful pulsed neutron source, allowing for better statistics and shorter measurement times. Research is currently underway to develop such sources~\cite{ref:starfire,KULCINSKI20161072}. 

\section{Acknowledgements}

This work was supported in part by Department of Energy Award No. DE-NA0003920, as part of the NNSA Consortium of Monitoring, Technology, and Verification (MTV). E.A. Klein gratefully acknowledges NNSA support for this research performed under appointment to the Nuclear Nonproliferation International Safeguards Fellowship Program sponsored by the Department of Energy, National Nuclear Security Administration’s Office of International Nuclear Safeguards (NA-241). 
The authors would like to thank their colleagues in MTV and the MIT Department of Nuclear Science and Engineering for encouragement and advice.

 \bibliography{apssamp}

\end{document}